\documentclass[aps,prb,reprint,superscriptaddress]{revtex4-2}

\usepackage{graphicx}% Include figure files
\usepackage{dcolumn}% Align table columns on decimal point
\usepackage{bm}% bold math
\usepackage{float}
\usepackage{subfigure}

\begin{document}

\title{Deep learning inter-atomic potential for irradiation damage in 3C-SiC}

\author{Yong Liu}
\altaffiliation{These authors contributed equally to this work}
\affiliation{State Key Laboratory of Nuclear Physics and Technology, School of Physics, CAPT, HEDPS, and IFSA, College of Engineering, Peking University, Beijing 100871, P. R. China}

\author{Hao Wang}
\altaffiliation{These authors contributed equally to this work}
\affiliation{Institute of Nuclear Physics and Chemistry, China Academy of Engineering Physics, Mianyang 621999, P. R. China}

\author{Linxin Guo}
\affiliation{State Key Laboratory of Nuclear Physics and Technology, School of Physics, CAPT, HEDPS, and IFSA, College of Engineering, Peking University, Beijing 100871, P. R. China}

\author{Zhanfeng Yan}
\affiliation{Institute of Nuclear Physics and Chemistry, China Academy of Engineering Physics, Mianyang 621999, P. R. China}

\author{Jian Zheng}
\affiliation{Institute of Nuclear Physics and Chemistry, China Academy of Engineering Physics, Mianyang 621999, P. R. China}

\author{Wei Zhou}
\affiliation{Institute of Nuclear Physics and Chemistry, China Academy of Engineering Physics, Mianyang 621999, P. R. China}

\author{Jianming Xue}
\email{Corresponding author: jmxue@pku.edu.cn}
\affiliation{State Key Laboratory of Nuclear Physics and Technology, School of Physics, CAPT, HEDPS, and IFSA, College of Engineering, Peking University, Beijing 100871, P. R. China}

\date{\today}

\begin{abstract}
    We developed and validated an accurate inter-atomic potential for molecular dynamics simulation in cubic silicon carbide (3C-SiC) using a deep learning framework combined with smooth Ziegler-Biersack-Littmark (ZBL) screened nuclear repulsion potential interpolation. Comparisons of multiple important properties were made between the deep-learning potential and existing analytical potentials which are most commonly used in molecular dynamics simulations of 3C-SiC. Not only for equilibrium properties but also for significant properties of radiation damage such as defect formation energies and threshold displacement energies, our deep-learning potential gave closer predictions to DFT criterion than analytical potentials. The deep-learning potential framework solved the long-standing dilemma that traditional empirical potentials currently applied in 3C-SiC radiation damage simulations gave large disparities with each other and were inconsistent with ab-initio calculations. A more realistic depiction of the primary irradiation damage process in 3C-SiC can be given and the accuracy of classical molecular dynamics simulation for cubic silicon carbide can be expected to the level of quantum mechanics.
\end{abstract}

\maketitle

\section{\label{sec:level1}Introduction}
Cubic silicon carbide has been widely used for electronic and nuclear applications due to its outstanding mechanical properties, high thermal conductivity, chemical stability, and good radiation response \cite{gao2004atomistic, oda2013study}. The mechanical and electrical properties of 3C-SiC are degraded due to the changes in microstructure when it subjects to high energy neutron in the nuclear environment. Understanding the primary irradiation process is of crucial importance to estimate the usable lifetime of this material.

Ab-initio molecular dynamics (AIMD) with density functional theory (DFT) and classical molecular dynamics (CMD) are the main tools to simulate the primary irradiation damage process at the atomic level beyond the limits of experimental techniques \cite{nordlund_historical_2019}. On the one hand, AIMD is accurate but computationally cost, which can only involve a few hundred atoms and several hundred picoseconds long \cite{zhang2018deep}. Many thousands of atoms can be knocked out of equilibrium position by one energetic ion or neutron generated from a nuclear reaction. Therefore, AIMD can not cover the atomic scale required to simulate the primary radiation damage process \cite{dudarev2013density}. On the other hand, CMD is efficient enough to satisfy the computational demand of primary irradiation dynamics simulation but the accuracy of simulation results greatly depends upon the employed inter-atomic potential. 

\begin{table}
	\caption{\label{table_potentials}The widely used empirical potentials for MD simulations of silicon carbide materials\cite{yan_molecular_2023}.}
\begin{ruledtabular}
	\begin{tabular}{llllll}
		&Potentials   &Applications\\
		\colrule
		&Tersoff\cite{tersoff1989modeling, halicioglu1995comparative, tersoff1994chemical, tersoff1990carbon, erhart2005analytical}              &Thermal properties, Mechanical properties,\\
		&                     &Electrical properties, Polishing,\\
		&                     &Ion implantation, Crystal growth,\\
		&                     &Irradiation damage, Amorphization,\\
		&                     &Fatigue damage, Shock damage\\
		&\\
		&Tersoff/ZBL\cite{devanathan1998displacement}       &Ion implantation, Irradiation damage\\	
		&\\
		&GW\cite{gao2002empirical}        &Irradiation damage\\
		&\\
		&GW/ZBL\cite{gao2001native}         &Crystal growth, Irradiation damage\\
		&\\
		&Vashishta\cite{vashishta_interaction_2007}            &Mechanical properties, Electrical properties,\\
		&                     &Deposition, Shock damage, Polishing\\
		&\\
		&MEAM\cite{huangts_molecular_nodate, kang2014governing}               &Crystal growth, Thermal properties,\\ &                     &Irradiation damage\\
		&\\
		&EDIP\cite{bazant_environment-dependent_1997, jiang_carbon_2012}              &Mechanical properties\\
	\end{tabular}
\end{ruledtabular}
\end{table}

The widely used potential functions for CMD simulations of silicon carbide materials and their applications were summarized in Table \ref{table_potentials}. Although the expression forms of different empirical analytic potentials are distinguished, the processes of their development are basically the same. First, a mathematical function based on a physical understanding of interatomic interactions in the material was proposed, with a handful of global fitting parameters. Then a series of labeled physical properties from experimental or ab-initio calculations were used to fit these adjustable parameters. Finally, this fixed expression will be used for predicting the energies and forces of the new configurations in MD simulations. Although all these potential functions in Table \ref{table_potentials} have taken the three-body effect and bond-angle effect into account to describe the many-body interaction in material and strong directionality of the covalent bonds, the true interactions in silicon carbide are determined by complex many-body interactions. The ability of traditional analytical force fields to fit the corresponding potential energy surface is inherently limited by their relatively simple functional forms and a few adjustable parameters. G. Lucas and L. Pizzagalli pointed out that the use of available empirical potentials is the largest source of errors to calculate threshold displacement energies in 3C-SiC and called for the improvement of existing potentials\cite{lucas2005comparison}. G.D.Samolyuk's study shows that the most commonly used Tersoff and MEAM potentials for SiC are inconsistent with the ab-initio calculation of defect energetics. Tersoff potential predicts a very high interstitial formation energy and high defect migration energy\cite{samolyuk2015molecular}. GW-ZBL potential gives a more realistic description of defect formation energy but still overestimates the defect migration energy barrier. \cite{samolyuk2015molecular}. Andrey Sarikov got divergent simulation results from different potentials (including Tersoff, Vashishta \cite{vashishta2007interaction} and an analytical bond order potential \cite{erhart2005analytical}) in their study of partial dislocations and stacking faults in 3C-SiC \cite{sarikov2019molecular}. The inaccurate depiction of these key physical quantities makes us lose confidence in the correctness of MD simulation results of radiation in 3C-SiC. A new potential that can accurately describe the inter-atomic interactions is urgently needed to be developed.

Recently, machine learning methods combined with DFT training data to build potential energy surfaces (PES) have been developed rapidly \cite{behler2011atom, zhang2018potential, behler2016perspective, botu2017machine, zhang2019active, byggmastar2019machine}. Compared with the construction method of traditional empirical potential, machine-learning potentials have a more powerful fitting ability due to their unintended preset expressions and abundant adjustable parameters \cite{zhang2018end}. Moreover, unlike the empirical potentials, which only fit a subset of properties, machine-learning potentials can sample configurations to train the PES as many as needed. Due to more general expressions and more complete training data, machine learning potentials can give a more accurate prediction of the PES to capture the underlying physical mechanism. A variety of CMD simulations with DFT accuracy in different areas have been carried out with the help of machine-learning potentials \cite{niu_ab_2020, zeng_complex_2020, zhang_phase_2021, wang_deep_2022}. In the field of radiation damage, a set of machine learning potentials have also been developed to simulate the irradiation damage processes for different materials such as fcc-aluminum\cite{wang2019deep}, tungsten \cite{byggmastar2019machine}, silicon\cite{hamedani_primary_2021}, and bcc-iron \cite{wang_machine-learning_2022}. So far, most of the machine learning potentials are for single-substance systems because the number of configurations needed to train the model increases exponentially with the increase of the number of principle elements.

In this work, we applied the DP-ZBL (Deep-learning Potential interpolated with ZBL) model \cite{wang2019deep} to train a deep-learning inter-atomic potential hybrid with ZBL screened nuclear repulsion potential for 3C-SiC. In order to capture the right pictures when atoms are extremely close to each other, which are frequent events happening in the irradiation damage process due to the high kinetic energy of atoms, the generally used ZBL screened nuclear repulsion potential \cite{ziegler1985stopping} has been interpolated into the deep learning framework so that short-range repulsion interaction between atoms can be accurately described. Here we refer Ref\cite{wang2019deep} for more details about the interpolation mechanism. Compared to the analytics empirical potential including Tersoff, MEAM, Vashishta, EIDP, and GW-ZBL, the DP-ZBL potential not only get the DFT accuracy for the near-equilibrium properties such as lattice constant, elastic properties, equation of state, and phonon dispersion but also give a correct description of short-range repulsion interaction. We put our concentration on correct prediction for defect formation energies and threshold displacement energies because these physical quantities play decisive roles in the irradiation process and there are large disparities between different studies using distinct inter-atomic potentials. The DP-ZBL model can terminate these controversies and get a more realistic molecular dynamics simulation of radiation damage in 3C-SiC.

\section{\label{sec:level1}Method}
\subsection{\label{sec:level2}Training process}
   The accuracy and transferability of DP models are determined by the quality of the training dataset. The training dataset should be complete enough to cover the target simulation space. To get a good description of the energy and force of dimer (Si-Si, Si-C, C-C), elastic properties, phonon dispersion, and defect formation energies, the corresponding configurations recorded in Table \ref{table_training_data} are taken as the initial training dataset.
  
  Data used to train the neural network were all generated by DFT calculations with VASP code \cite{kresse1993ab,kresse1994ab,kresse1996efficiency}. General gradient approximate (GGA) with PBE \cite{perdew1996generalized} exchange-correlation functional has been used. The plane-wave cutoff energy was set high enough to 600eV to cover large deformation in the irradiation process. Consistent spacing between k-points in Brillouin zone (KSPACING = 0.15 $\rm{\AA^{-1}}$) was integrated using Gamma centered grid for all configurations. Gaussian smearing with 0.03 eV width was applied to help convergence. Spin-polarized calculations were considered to account for the possible spin polarization of various defect configurations.

  \begin{table}
  	\caption{\label{table_training_data}Configurations included in the initial training database for corresponding properties.}
  	\begin{ruledtabular}
  		\begin{tabular}{llllll}
  			&Concerned properties   &Configuration type\\
  			\colrule
  			&Bulk properties           &Equilibrium state\\
  			&                          &Compressed\\
  			&                          &Stretched\\
  			&\\
  		    &Thermo properties         &Atom displaced\\
  		    &\\
  		    &Elastic properties        &Elastically distorted\\
  		    &\\
  		    &Defect properties         &Vacancy with strain\\
  		    &                          &Antisite with strain\\
  		    &                          &Tetrahedral interstitial with strain\\
  		    &\\
  		    &Liquid phase              &Frames of liquid trajectory\\
  		    &\\
  		    &Short-range interactions  &Dimers\\
  		    &\\
  		    &Irradiation damage        &Frames of PKA activation trajectory
  		\end{tabular}
  	\end{ruledtabular}
  \end{table}

After the initial training dataset was used to kick off the first round train, an active learning training process was performed by DPGEN\cite{zhang_dp-gen_2020} to sample more configurations into the training dataset.The active learning process terminated when we validated that the potential energy is good enough. In this work, we went through fifty active learning iterations to get the final potential energy model, which made the training dataset sampling 33898 configurations in total. The temperature of NVT setting in the exploration stage went up from 300K to 4000K and the environmental pressure is set to one atmosphere.

\subsection{\label{sec:level2}Interpolation for short range repulsion} 
  We refer to our previous work Ref\cite{wang2019deep} to get the details of the interpolation method between DP and ZBL. Meanwhile, R.E.Stoller's work in Ref\cite{stoller_impact_2016} provides a more systematic and effective procedure to bridge the equilibrium and short-range parts, which can avoid lots of invalid tests.  First, the table recording the energy of dimer Si-Si, Si-C, C-C at a short range ($\rm{0.001\AA, 1.200 \AA}$) with $\rm{0.001 \AA}$ step was generated by ZBL formula. The dimer configurations of Si-Si, Si-C, and C-C in range ($\rm{0.5, 5.0}$) with $\rm{0.05 \AA}$ step were calculated by DFT and added to the training dataset. Then the DP and the ZBL potential were smoothly docked in the interval ($\rm{1.0 \AA, 1.2\AA}$) to let ZBL plays its role in the short range and the DP model dominates at equilibrium condition.
  
\begin{figure}[!t]
  	\includegraphics[width=0.9\columnwidth]{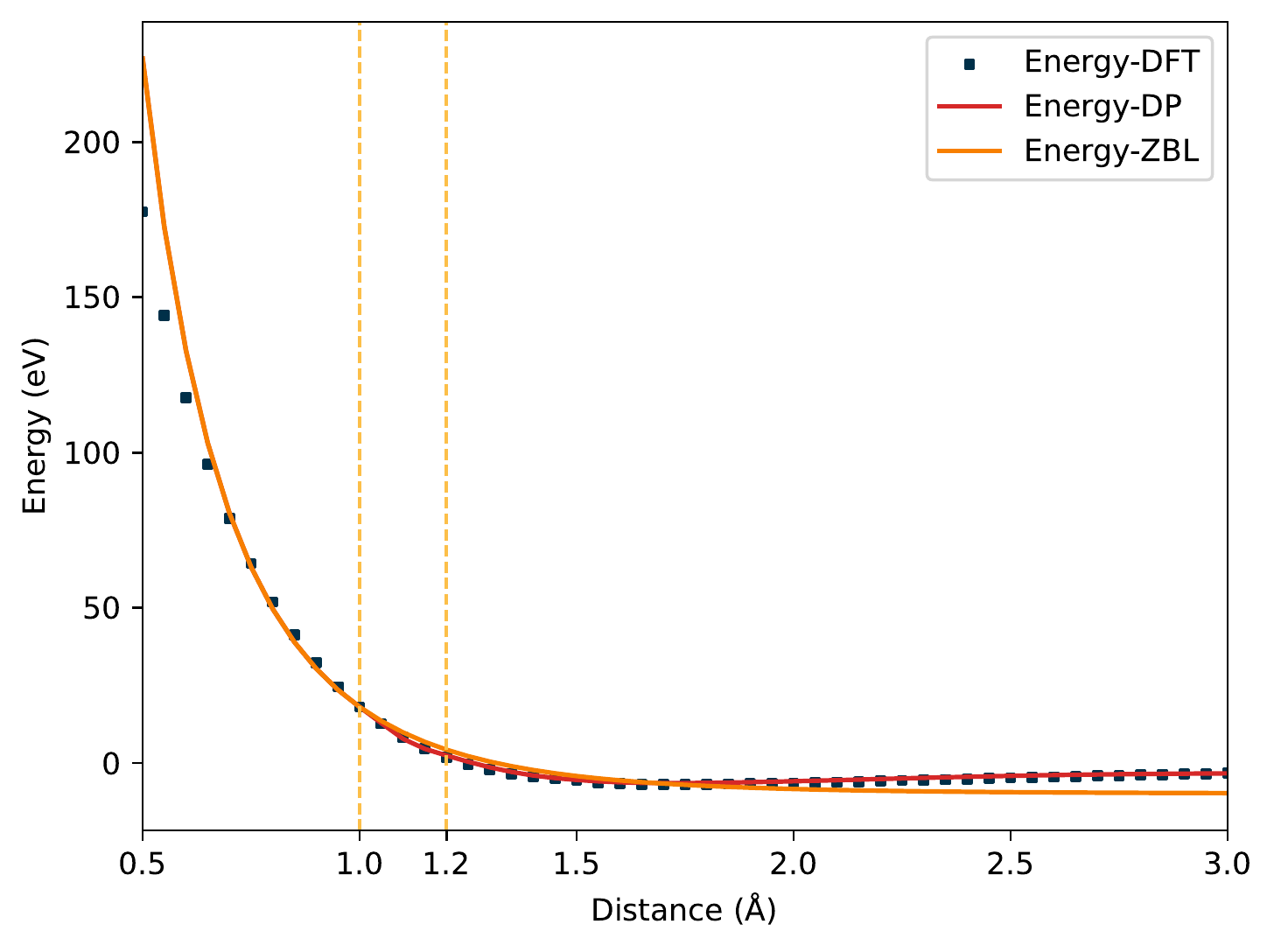}
  	\caption{\label{label-energy-of-dimer}The calculated energy versus distance curve of Si-C dimer by DFT, ZBL and DP-ZBL.}
  \end{figure}

  As shown in Figure \ref{label-energy-of-dimer}, when Si-C dimer's distance is less than 1.0 $\rm{\AA}$, DP-ZBL is consistent with pure ZBL potential and when Si-C dimer's distance is larger than 1.2 $\rm{\AA}$, DP-ZBL gives results dominated by DFT calculation. And for the Si-C dimer's distance between 1.0 and 1.2 $\rm{\AA}$, DP-ZBL has smoothly switched from ZBL to DFT.
  
\section{\label{sec:level1}Result}
\begin{figure}[!t]
	\includegraphics[width=1.0\columnwidth]{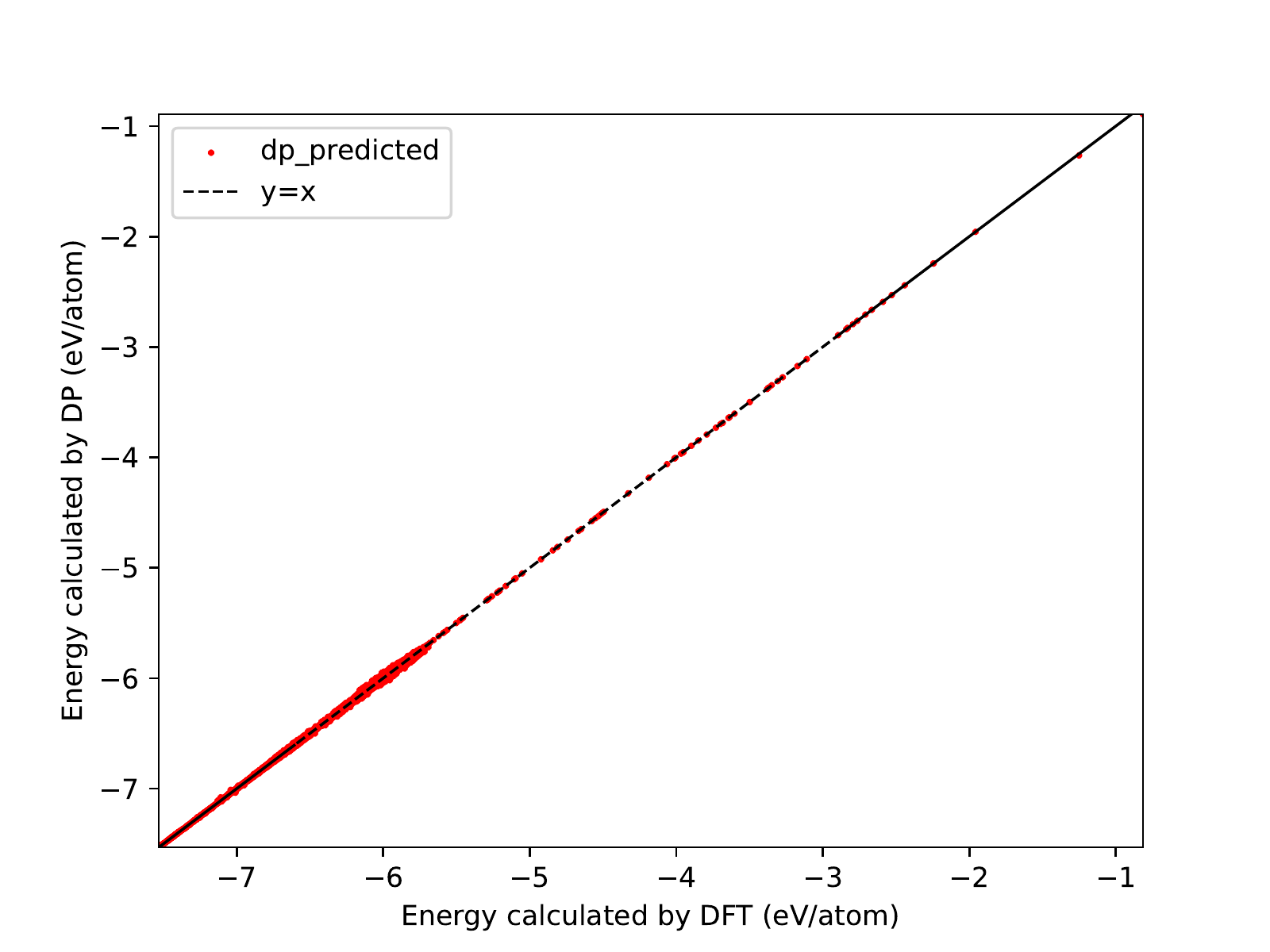}
	\caption{\label{label-energy-prediction}Comparison of energy prediction by DFT and DP for the final training set. Both axes represent the energy of the configuration divided by the number of total atoms in the configuration.}
\end{figure}

\begin{figure}[!t]
	\includegraphics[width=1.0\columnwidth]{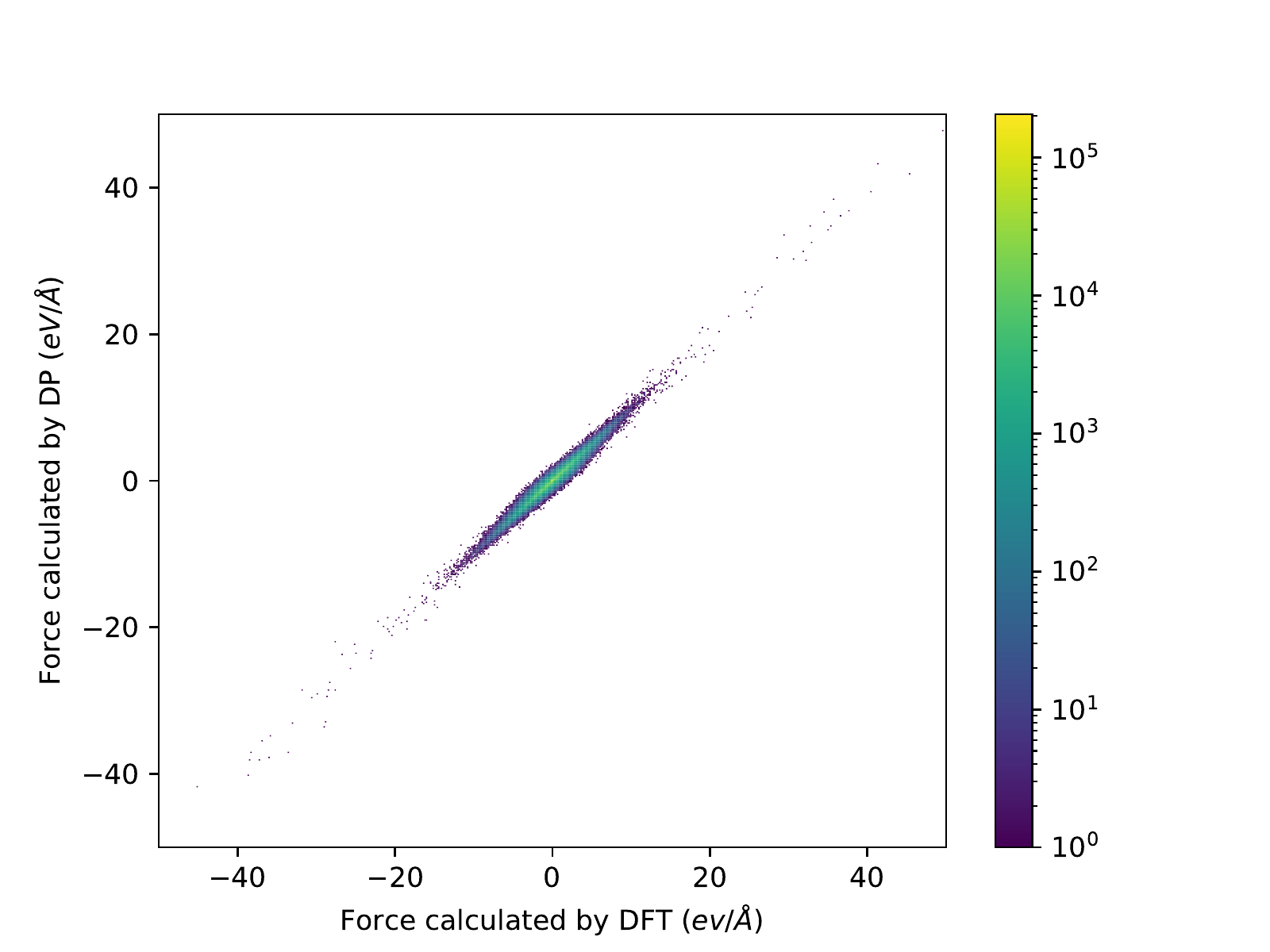}
	\caption{\label{label-force-prediction}Comparison of force prediction by DFT and DP for the final training set. The color bar indicates the density of data. Force greater than 50 eV/$\rm{\AA}$ are not shown in the figure.}
\end{figure}

We compared the energies calculated by the DP-ZBL model and the DFT method for all the configurations in the training dataset. As shown in Fig. \ref{label-energy-prediction} and Fig. \ref{label-force-prediction}, DP-ZBL prediction is consistent with DFT calculation as the points basically distribute around the y=x reference line. The root-mean-squared-errors (RMSEs) of the energies and the forces are 0.01 $\rm{meV/atom}$ and 0.16 $\rm{eV/\AA}$ respectively, which are within the accuracy allowed for typical Deepmd-kit training \cite{zeng2020exploring} compared with the range of energy and force.

Several near-equilibrium material properties have been calculated by the DP-ZBL and the DFT method and four different empirical potentials including Tersoff-ZBL, GW-ZBL, MEAM, and Vashishta for comparison, as shown in Table \ref{bulk_properties}. including lattice parameter $\rm{a_0}$, elastic constants $\rm{C_{11}}$, $\rm{C_{12}}$, $\rm{C_{44}}$, bulk modulus $\rm{K_H}$, Young's modulus $\rm{E_H}$, shear modulus $\rm{G_H}$ (all the moduli are in Hill noting), cohesive energy $\rm{E_{coh}}$, and defect formation energy. All results were in excellent agreement with DFT data computed in this work or referred from other works. In contrast, most empirical potentials do not give good predictions for most properties. For instance, the Tersoff-ZBL potential underestimates the lattice constant which is the most basic physical quantity in simulations. However, Tersoff-ZBL works well with the elastic contants including $\rm{C_{11}}$, $\rm{C_{12}}$, $\rm{C_{44}}$. The GW-ZBL potential almost mispredicts the elastic response relationship. Meanwhile, the MEAM potential has a slight error and the Vashishta potential underestimates the $\rm{C_{44}}$ constant. 

The formation energy of defects in 3C-SiC is defined as follows:
\begin{equation}
	\rm{E_{f}= E_{defect}-E_{perfect} + n_{Si}\mu_{Si} + n_{C} \mu_{C}}
\end{equation}

\noindent where $\rm{E_{perfect}}$ and $\rm{E_{defect}}$ are the total energy of a perfect 3C-SiC supercell and a supercell containing a defect, respectively. The integer $\rm{n_{Si}}$ gives the number of Si atoms removed from ($\rm{n_{Si} \textgreater 0}$) or add to ($\rm{n_{Si} \textless 0}$) the perfect supercell, and $\rm{n_C}$ follows the same logic. The $\rm{\mu_{Si}}$ and $\rm{\mu_{C}}$ are respectively the chemical potential of the Si atom and the C atom in the 3C-SiC bulk environment. In this work, all defect formation energies were calculated for the Si-rich condition, which means the chemical potential of the Si atom in 3C-SiC is limited to that in the cubic silicon crystal. In this context, $\rm{\mu_{Si} = \mu_{Si}(bulk)}$ and $\rm{\mu_{C} = \mu_{SiC} - \mu_{Si}}$, where $\rm{\mu_{SiC}}$ is the chemical potential of Si-C atom pair in 3C-SiC \cite{chen2018ab}. The results calculated by the DP-ZBL model match well with the DFT results for all the defect configurations in this work. The GW-ZBL potential underestimates the defect formation energies of most configurations except the antisite of C replacing Si. The Tersoff-ZBL potential underestimates the defect formation energies of vacancies and antisites but overestimates the defect formation energies of the tetrahedral interstitial. Defect formation energy is a key physical quantity that reflects the accuracy of irradiation simulation. To sum up, the DP-ZBL model performs much better than the four listed empirical potentials for the predictions of the near-equilibrium properties.

\begin{table*}
	\caption{\label{bulk_properties} Basic properties of 3C-SiC: lattice constant $\rm{a}$, cohesive energy $\rm{E_{coh}}$, elastic constants $C_{\rm{11}}$, $C_{\rm{12}}$ and $C_{\rm{44}}$, bulk modulus $K_{\rm{H}}$ (Hill), Young's modulus $E_{\rm{H}}$ (Hill), shear modulus $G_{\rm{H}}$ (Hill), Poisson's ratio $\rm{v_H}$(Hill).} 
	\begin{ruledtabular}
		\begin{tabular}{cccccccccc}
			& Properties   &$\rm{DFT_{Ref}}$    &$\rm{DFT_{Current}}$    &DP-ZBL  &Tersoff-ZBL &GW-ZBL &MEAM &Vashishta &EDIP\\
			\colrule
			&$\rm{a_0 (\AA)}$  &4.3805\footnote{Reference \cite{pizzagalli_accurate_2021}}              &4.3784                   &4.3778  &4.2796      &4.3600 &4.3595  &4.3582  &4.3624\\
		    &$\rm{E_{coh}}$      &-15.0624\footnotemark[1]    &-15.0643     &-15.0630      &-12.68   &-12.82  &-12.86 &-12.68  &-12.67\\
			&$\rm{C_{11}(GPa)}$&383.9\footnotemark[1]               &380.8                    &375.3   &445.7       &265.2  &396.5   &390.1 &396.8\\
			&$\rm{C_{12}(GPa)}$&127.6\footnotemark[1]               &126.8                    &124.0   &138.7       &219.3  &147.1   &142.8 &140.5\\
			&$\rm{C_{44}(GPa)}$&239.5\footnotemark[1]               &240.1                    &223.1   &220.0       &101.1  &135.6   &136.9 &170.3\\
			&$\rm{K_H (GPa)}$  &213.0\footnotemark[1]              &211.5                   &207.7  &241.0      &234.6 &230.3  &225.2 &226.0\\
			&$\rm{E_H (GPa)}$  &432.8\footnotemark[1]              &431.4                  &413.9   &452.2      &156.5 &330.6  &330.1 &372.5\\
			&$\rm{G_H (GPa)}$  &186.3\footnotemark[1]              &186.0                  &177.2  &190.4     &56.4  &131.1  &131.4 &152.0\\
			&$\rm{v_H}$        &0.16\footnotemark[1]               &0.16                     &0.17    &0.19      &0.39   &0.26    &0.26 &0.23\\
			%&$\rm{T_m} (K)$   &$3130^{\rm{a}}$        &3272     &   &5600  &3800 &4600   &3400  \\
			&$\rm{V_{Si}}$         &7.75\footnote{Reference \cite{hu_thermodynamic_2014}}           &7.72   &7.66   &8.24  &6.89   &4.90  &12.73   &4.60\\
			&$\rm{V_C}$            &4.09\footnotemark[2]           &4.21   &4.10   &3.76  &-0.84  &1.06  &-3.38   &1.22\\
			&$\rm{C_{Si}}$         &3.94\footnotemark[2]           &3.92   &3.91   &3.29  &8.87   &2.74  &33.48   &3.02\\
			&$\rm{Si_C}$           &3.29\footnotemark[2]           &3.35   &3.31  &4.90  &0.74   &3.84  &-3.32   &2.04\\
			&$\rm{Si_{TSi}}$       &10.87\footnote{Reference \cite{sun_interaction_2017}}          &10.22  &10.26  &16.65 &3.23   &4.00  &-2.07   &11.69\\
			&$\rm{Si_{TC}}$        &9.04\footnotemark[2]           &8.47   &8.34   &16.48 &0.33   &3.22 &-3.41   &12.24\\
			&$\rm{C_{TSi}}$        &10.09\footnotemark[3]          &9.97   &9.88   &4.89  &7.86   &9.08  &17.84   &6.69\\
			&$\rm{C_{TC}}$         &11.10\footnotemark[2]          &10.96  &10.86  &7.89  &8.22   &3.05  &21.16   &8.29\\
		\end{tabular}
	\end{ruledtabular}
\end{table*}

\begin{figure}[!t]
	\includegraphics[width=0.9\columnwidth]{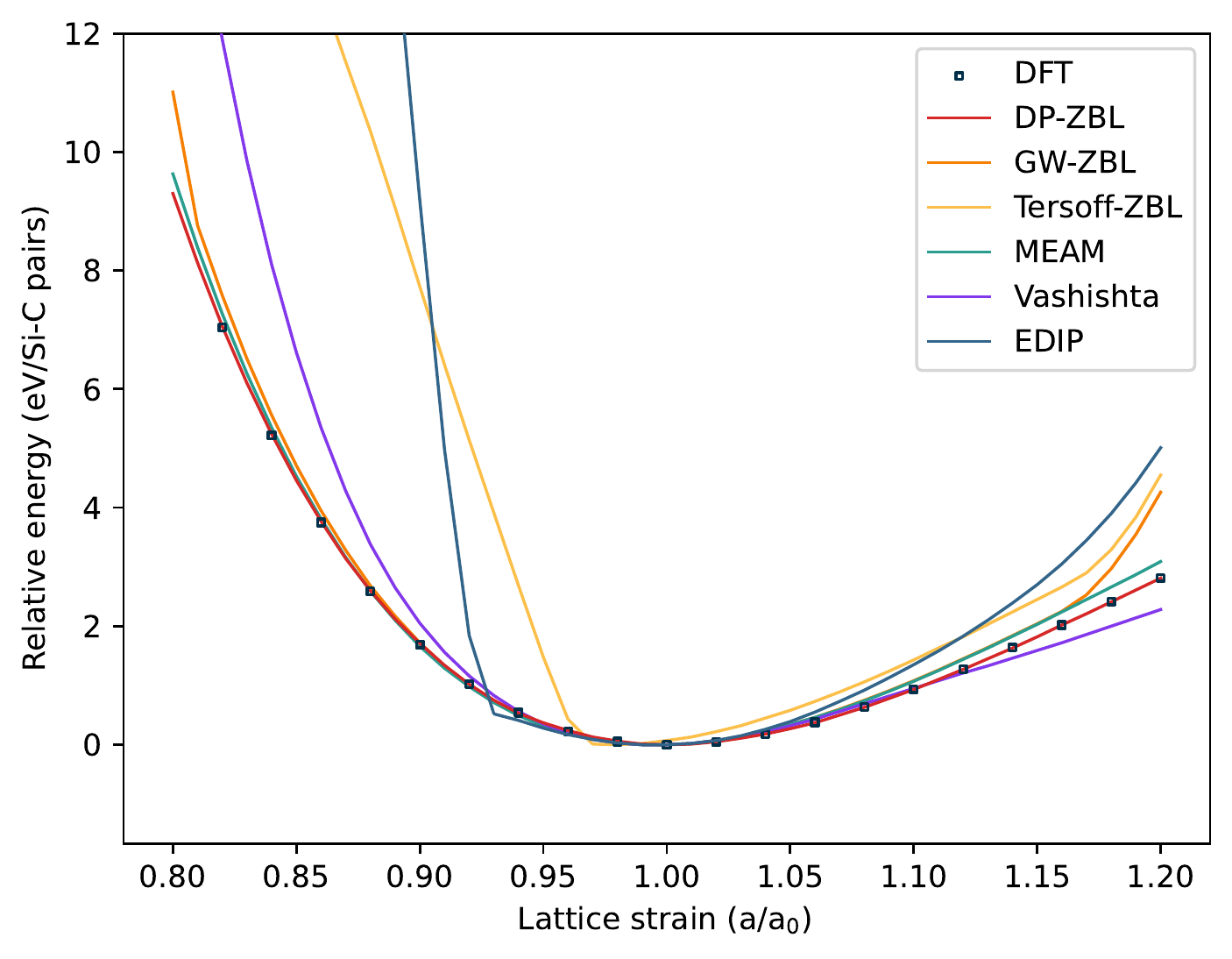}
	\caption{\label{label-equation-of-state-dp}Equation of state of the 3C-SiC  phase as computed by different empirical potentials and the DP-ZBL model and the DFT method.}
\end{figure}

The equation of state curves of the 3C-SiC phase computed by different empirical potentials, the DP-ZBL model and the DFT method are illustrated in Figure \ref{label-equation-of-state-dp}. The DP-ZBL model well reproduces the DFT results, which indicates that the DP-ZBL potential is capable to cover the high compressing and stretching conditions. By contrast, the Tersoff-ZBL and Vashishta potentials produce large errors compared with the DFT criterion in the compressing condition. The GW-ZBL and Tersoff-ZBL potentials overestimate the potential energy when the system is stretched to 1.2 times relative to the equilibrium state. 

\begin{figure}[!t]
	\includegraphics[width=0.9\columnwidth]{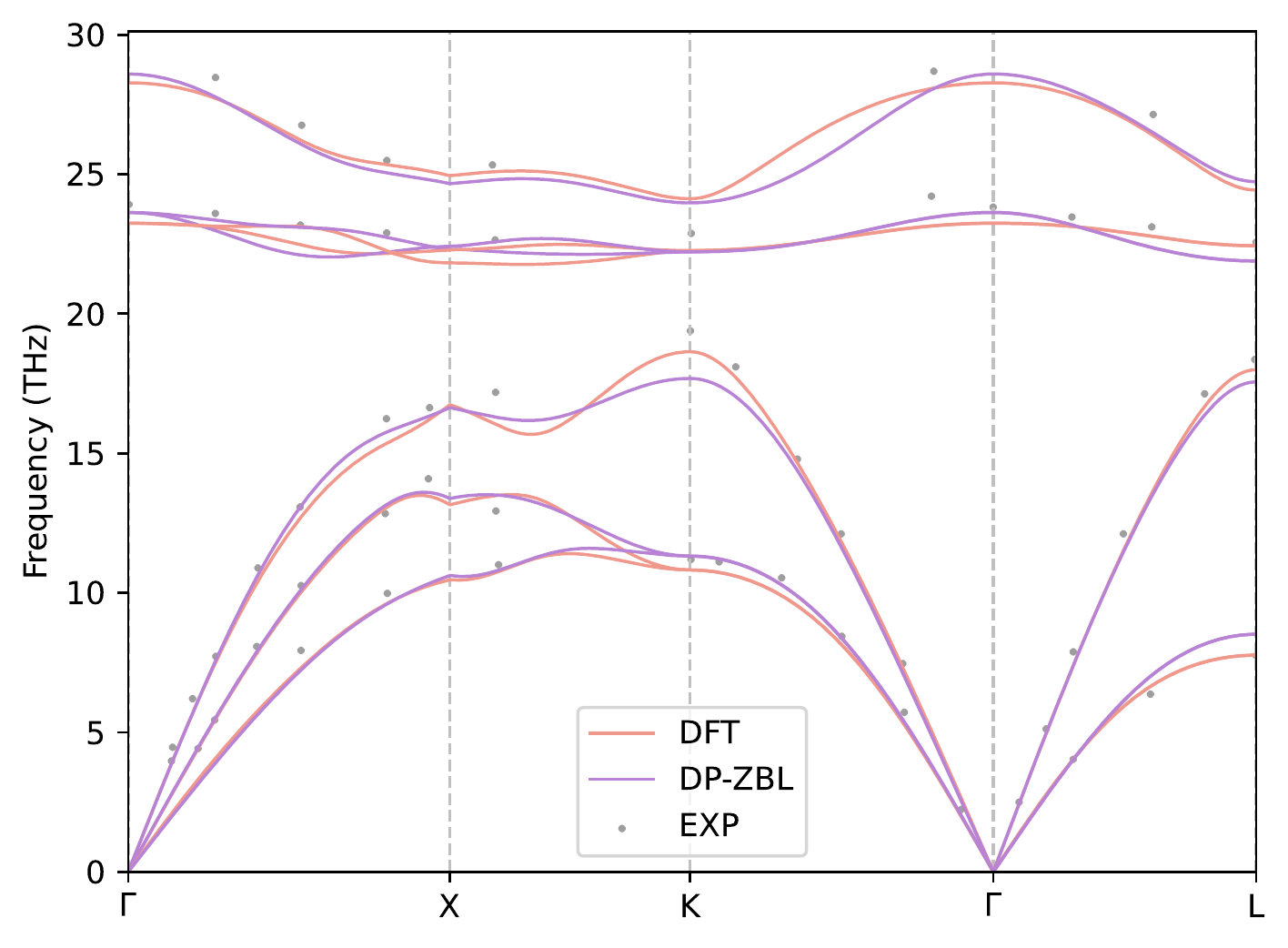}
	\caption{\label{label-phonon-dispersion-dp}Phonon dispersion curve of 3C-SiC calculated by our DP-ZBL model and the DFT method, as well as the experimental data measured by Serrano et al. \cite{serrano_determination_2002} using IXS.}
\end{figure}

The phonon dispersion curve of the 3C-SiC phase has been calculated along the high symmetry directions of $\rm{\Gamma}$-X-K-$\rm{\Gamma}$-L by our DP-ZBL model and the DFT method. The force constants were calculated by density functional perturbation theory using VASP for the DFT method and were calculated using PhonoLammps for the DP-ZBL model. Then the Phonopy package \cite{togo_first_2015} was used to compute the phonon dispersion relations. Non-metallic crystals are polarized by atomic displacements and the generated macroscopic field changes force constants near $\rm{\Gamma}$ point\cite{pick1970microscopic}. To take this into consideration, phonon frequencies at general q-points with long-range dipole-dipole interaction were calculated by the method of Gonze et al. \cite{gonze1994interatomic, gonze1997dynamical}. The Bron effective charges ($\rm{Z^*)}$ and dieletric constant ($\epsilon_0$) calculated by GGA functional ($\rm{Z^{*}_{Si}}=\rm{Z^{*}_{C}}=2.69$, $\epsilon_0=6.99$) are in agreement with the experimental value ($\rm{Z^{*}_{Si}}=\rm{Z^{*}_{C}}=2.69$ \cite{olego_pressure_1982}, $\epsilon_0=6.52$ \cite{madelung1982physics}). As shown in Fig. \ref{label-phonon-dispersion-dp}, both acoustic branches and optical branches of the six phonon modes generated by the DP-ZBL model are close to the DFT results. In addition, the results from both theoretical models match well with the experimental data measured by Serrano et al. \cite{serrano_determination_2002} at room temperature using inelastic x-ray scattering (IXS) with the synchrotron radiation source, which means a good description of crystal thermal response of 3C-SiC can be predicted with our DP-ZBL potential. 

Threshold displacement energy ($\rm{E_d}$) is defined as the minimum kinetic energy transferred to a lattice atom to displace it from its original Wigner-Seitz cell and form a stable Frenkel pair \cite{nordlund_molecular_2006}. $\rm{E_d}$ is a critical physical parameter for estimating damage production and predicting the defect profile under ion, neutron, and electron irradiation \cite{wang2013ab}. For example, $\rm{E_d}$ is a key input in large-scale irradiation simulation packages such as SRIM and TRIM to determine implantation profiles in doping processes or calculate damage accumulation in materials \cite{lucas_ab_nodate}. In this work, the $\rm{E_d}$ for both Si and C primary knock-on atoms (PKAs) along four typical low-index crystallographic directions including [100], [110], [111], and $\rm{[\overline{111}]}$ were calculated using different interatomic potentials for comparison. The simulations were performed at 300 K. A noncubic simulation box of $\rm{10 \times 10 \times 12}$ supercell (9600 atoms) with periodic boundary conditions was used. Kinetic energies in 0.5 eV increments were progressively assigned to a specific PKA atom in the central area to find the minimum energy. The simulation system was relaxed in the canonical ensemble (NVT) for 10 ps at 300K followed by cascade simulation in the microcanonical ensemble (NVE) for 10 ps. The Wigner-Seitz defect method was used to identify defects. From the calculation results summarized in Table \ref{table_threshold_displacement_energy}, the ${\rm{E_d}}$ values generated by our DP-ZBL are close to the DFT calculations carried out by Zhao et al\cite{zhao_influence_2012}. GW-ZBL, MEAM, and Vashishta show divergence from the DFT calculation in multiple crystal directions. Tersoff performs better than the three other empirical potentials, but it is also out of line with the DFT values for Si PKA in direction $\rm{[\overline{111}]}$ and C PKA in direction [110]. After the comparison, it is clear that the DP-ZBL potential yield better $\rm{E_d}$ values than the empirical potential functions.

\begin{table*}
	\caption{\label{table_threshold_displacement_energy} Threshold displacement energy calculated by DFT, DP-ZBL, and a range of empirical interatomic potentials.}
	\begin{ruledtabular}
		\begin{tabular}{cccccccccc}
			&   &DFT\cite{zhao_influence_2012} &DP-ZBL &Tersoff-ZBL &GW-ZBL &MEAM  &Vashishata  &EDIP\\
			\colrule
			&Si $\rm{[100]}$             &41      &33.5    &47.0   &20.5   &36.5   &29.5   &42.0\\
			&Si $\rm{[110]}$             &50      &47.0    &41.0   &23.5   &26.5   &22.5   &42.0\\
			&Si $\rm{[111]}$             &21      &23.0    &26.0   &12.5   &25.5   &46.0   &21.5\\
			&Si $\rm{[\overline{111}]}$  &33      &43.5    &44.0   &31.5   &26.0   &35.0   &22.5\\
			&C $\rm{[100]}$              &18      &15.0    &15.5   &11.5   &23.0   &47.5   &15.5\\
			&C $\rm{[110]}$              &19      &17.0    &29.0   &15.5   &27.0   &47.5   &15.5\\
			&C $\rm{[111]}$              &17      &15.5    &23.5   &8.0    &19.5   &150.5  &15.0\\
			&C $\rm{[\overline{111}]}$   &50      &43.5    &53.0   &23.0   &33.5   &15.5   &16.0\\
		\end{tabular}
	\end{ruledtabular}
\end{table*}

\begin{figure}[!t]
\includegraphics[width=0.9\columnwidth]{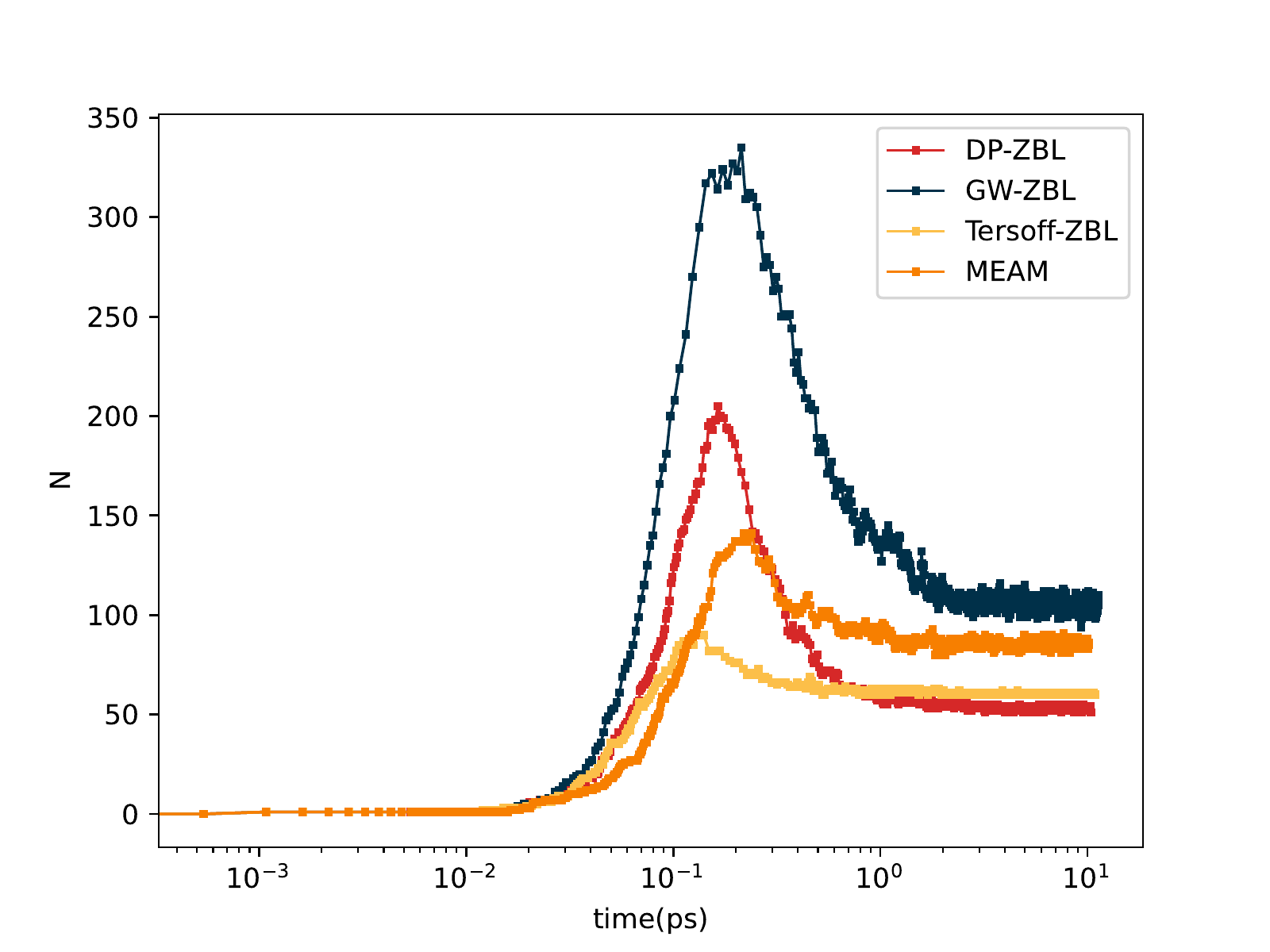}
\caption{\label{label-vacancy-comparison}Comparison of the production of vacancies time dependent relationship caused by a single 5 keV Si PKA.}
\end{figure}

\begin{figure}[!t]
	\includegraphics[width=0.9\columnwidth]{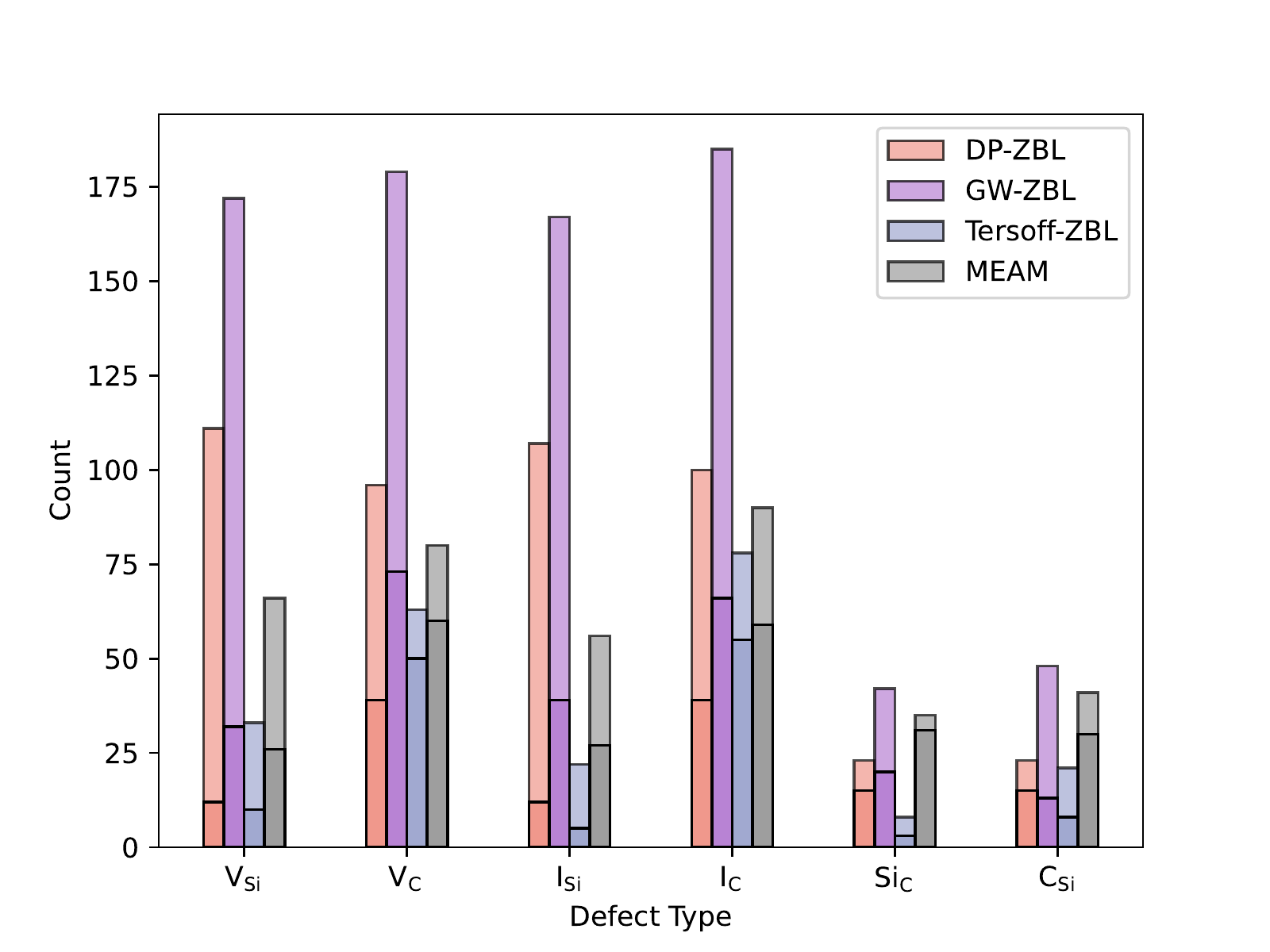}
	\caption{\label{label-max-and-stable-defects-comparison}Peak (slightly transparent) and stable numbers (solid) of defects caused by a single 5 keV Si PKA.}
\end{figure}

Then we carried out cascade simulations involving $60 \times 60 \times 60$ unit cells (1728000 atoms) using different potentials including Tersoff-ZBL, MEAM, GW-ZBL, and our DP-ZBL, which are widely used in the simulations of irradiation damage. The simulated system was equilibrated for 10 ps with timesteps of 1 fs in NVT ensemble at 300 K. Then a single Si atom in the central area was given kinetic energy of 5.0 keV in $\rm{[135]}$ direction to initialize the cascade while holding zero total momentum of the system. The cascade evolved for 10 ps in the NVE ensemble and during this period the timesteps were modified in order that the distance traveled by the fastest particle in the system was less than 0.1 $\rm{\AA}$ per timestep. To dissipate the heat generated by the PKA, the NVT ensemble of 300K was applied to the boundary region (2 times the lattice constant, about 8.8 $\rm{\AA}$). The Wigner-Seitz cell analysis method was used to determine the defects number with Ovito software \cite{stukowski2009visualization}. The time-dependent curves of defect amount are shown in Figure \ref{label-vacancy-comparison}. The peak of defects amount with the GW-ZBL potential is much higher than the results calculated by the other three potentials and so is the number of residual defects. We infer this is because the GW-ZBL potential underestimates the threshold displacement energy of either silicon or carbon atom. There are more than 65\% recombine of interstitials and vacancies during the annealing process with the GW-ZBL potential and the DP-ZBL potential, while only 20\% $\rm{\sim}$ 30\% recombine with the two other potentials. Defect yields and ratios predicted by the four potentials are distinguishing. Among the different point defects, vacancies and interstitials of carbon and silicon are dominant whether in thermal peak or stable state as shown in Figure \ref{label-max-and-stable-defects-comparison}. The DP-ZBL model predicts slightly more vacancies and interstitials of silicon than carbon at the thermal peak and residual vacancies and interstitials of carbon but domains after annealing. 

\section{\label{sec:level1}Conclusions}
In this work, a potential energy surface for silicon carbide was developed with our DP-ZBL model. Compared with the four most commonly used empirical interatomic potentials for SiC, the DP-ZBL potential can not only give a better performance on the prediction of near-equilibrium properties including lattice constant, elastic coefficients, equation of state, phonon dispersion, and defect formation energies but also depict a more precise picture of irradiation damage. More accurate values of key parameters in irradiation such as threshold displacement energy and defect migration energy can be gotten by using the DP-ZBL potential. Furthermore, our work provides a feasible approach to figuring out the primary irradiation damage process in covalent compound materials with ab-initio accuracy. 

\begin{acknowledgments}
This work is supported by National Natural Science Foundation of China (Grants No.12135002, No.12205269 and No.U20B2010), fund of Science and Technology on Plasma Physics Laboratory (No.22ZZJJ0601) and the Nuclear Energy Development Project. We are grateful for computing resources provided by High performance Computing Platform of Peking University.
\end{acknowledgments}

\bibliography{SiC-Deep-Potential-Manuscript}

\end{document}